\documentclass[aps,graphicx,prl,twocolumn,superscriptaddress]{revtex4}
\usepackage[dvips]{graphics}
\usepackage[dvips]{graphicx}

\begin{document}
\title{Charge order versus Zener polarons: ferroelectricity in manganites.}
\author{Dmitry V.~Efremov}
\affiliation{%
Laboratory of Solid State Physics, Material Science Center,
University of Groningen, Nijenborgh 4, 9747 AG Groningen, The
Netherlands }
\author{Jeroen van den Brink}
\affiliation{%
Institut-Lorentz for Theoretical Physics, \\  Universiteit Leiden,
P.O.Box 9506, 2300 RA Leiden, The Netherlands }
\author{Daniel I.~Khomskii}
\affiliation{%
Laboratory of Solid State Physics, Material Science Center,
University of Groningen, Nijenborgh 4, 9747 AG Groningen, The
Netherlands }
\affiliation{%
Universit\"at zu K\"oln, Zuelpicher Str. 77, 50937 K\"oln, Germany
}
\date{\today}
\begin{abstract}
We show that in manganites close to half-doping novel non-bipartite magnetic
phases appear due to the interplay between double exchange, superexchange
and orbital ordering. In considerable part of the phase diagram the groundstate
has a magnetic order that is intermediate between the canonical magnetic CE-phase
and a state that we identify as the recently observed Zener polaron state. The
intermediate phase shows a type of charge ordering that breaks inversion symmetry
and is therefore predicted to be ferroelectric.
\end{abstract}
\maketitle

{\it Introduction.} The wide variety of phenomena that are
observed in transition metal (TM) oxides is caused by the
interplay of various degrees of freedom. An intrinsic property of
doped TM oxides is the formation of superstructures that are
related to the charge, spin, lattice and orbital degrees of
freedom. Different types of charge ordering (CO), orbital ordering
and magnetic ordering may exist depending on e.g. doping
concentration, temperature and pressure. Charge order is usually
considered to be an ordering of TM ions with different valence,
for instance the alternation of Mn$^{3+/4+}$ in
La$_{0.5}$Ca$_{0.5}$MnO$_{3}$ or Fe$^{2+/3+}$ in Fe$_{3}$O$_{4}$.
Such a CO state we refer to as a site-centered (or TM-centered)
CO, or equivalently, a site-centered charge density wave (SCDW).
There is however another possibility for charge ordering
---the formation of a bond-centered charge density wave
(BCDW). The BCDW is very similar to the state that is observed in
Peierls distorted quasi one-dimensional materials. In a TM oxide
the BCDW  is actually an oxygen-centered charge ordered state as
oxygen ions are located on the TM-TM bonds in the typical (e.g.
perovskite) crystal structures of these systems. The possibility
to have such an oxygen-centered BCDW was studied theoretically for
the compounds RNiO$_3$ (with R =Pr, Nd, etc.) \cite{Mizokawa}.

For about the last 50 years it was presumed that in manganites of
the type R$_{1-x}$Ca$_x$MnO$_{3}$ (with R = La,  Pr, etc.) only
the Mn-centered SCDW occurs. At e.g. $x=1/2$ this SCDW state has
the well-known orbital ordered CE-type magnetic structure, see
Fig.1a, an ordering for which there is abundant experimental and
theoretical support \cite{Wollan,BKK}. However, Daoud-Aladine {\it
et al.} challenge this canonical picture in their recent study
\cite{Carvajal}. These authors conclude that in single crystalline
Pr$_{0.6}$Ca$_{0.4}$MnO$_{3}$ another type of superstructure is
more consistent with the experimental data. This superstructure is
a BCDW, which they refer to as the Zener polaron state (Fig.1b).
This state is made out of dimers in which all Mn ions have the
same valence (all Mn$^{3+x}$), quite different from the canonical
SCDW  that holds Mn ions with unequal valence. Recently the BCDW
state was also found theoretically in Hartee-Fock calculations of
the electronic structure \cite{Patterson,Littlewood}. The presence
of a BCDW  amounts to a drastic change in our understanding of
charge ordering ---not only for manganites. It may imply that in
all cases of charge order in oxides one should have in mind,
besides the usual site-centered CO, alternative states with
bond-centered superstructures.

\begin{figure}
\includegraphics[width=0.32\columnwidth]{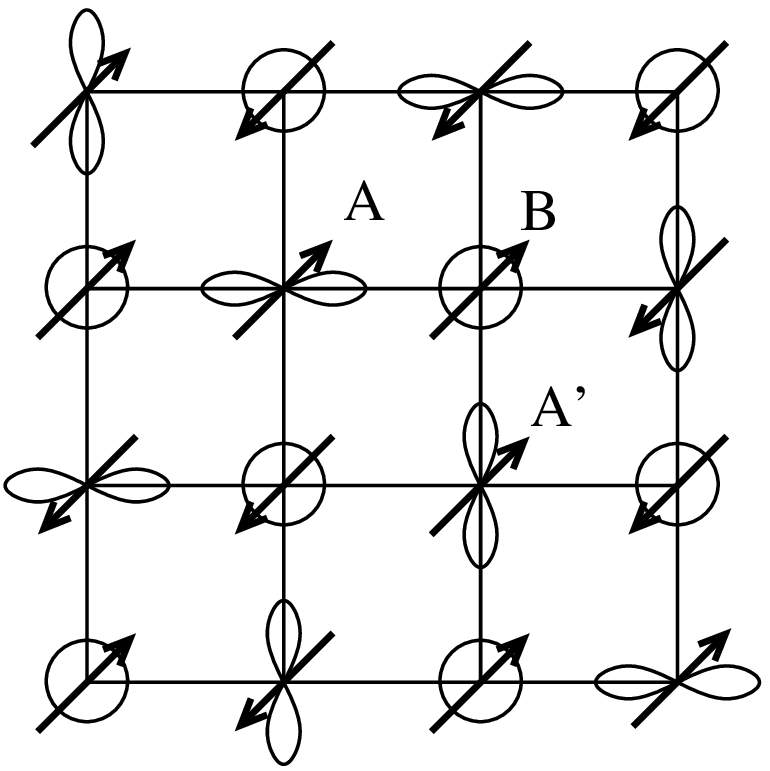}
\includegraphics[width=0.32\columnwidth]{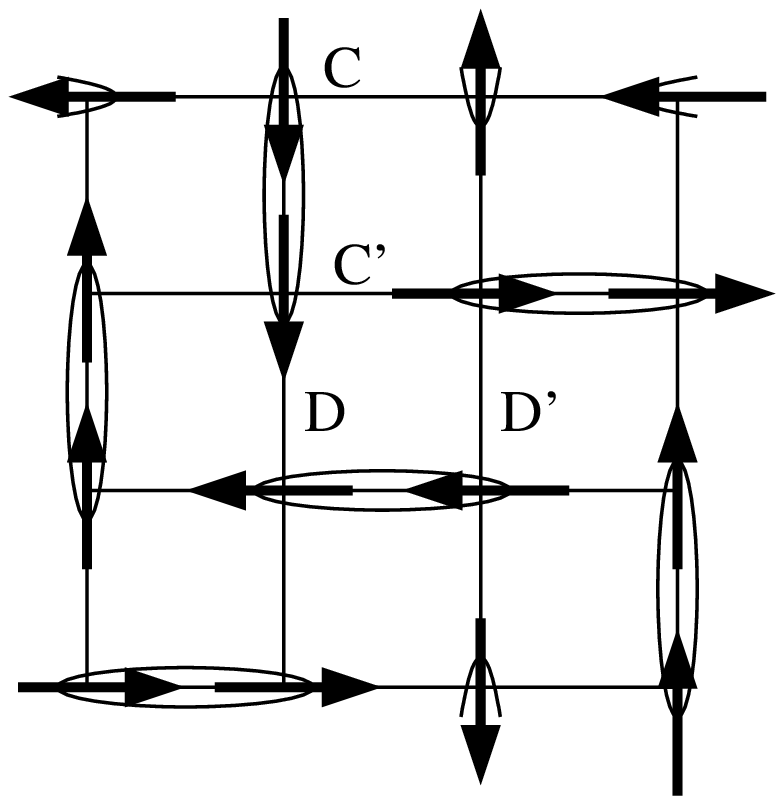}
\includegraphics[width=0.3\columnwidth]{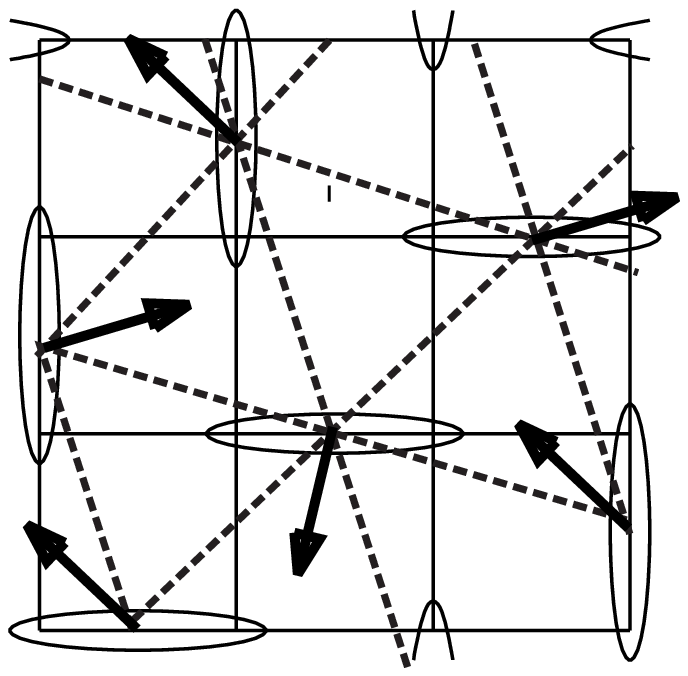}
\caption{\label{fig1} Three types of magnetic/orbital
superstructures: a) CE phase; b) Orthogonal ($\perp$) phase; c)
120$^{\circ}$ - phase. Directions of spins are shown by arrows.}
\end{figure}

In this paper we consider the regions of stability of  different
superstructures in doped manganites close to half-doping, and show
that indeed the bond-centered charge density wave (Zener polaron)
state may be favorable in part of the phase diagram.  Our results
reconcile the conflicting conclusions of Refs. \cite{Wollan} and
\cite{Carvajal}: the conventional CE state is still better at
exactly half-doping $x = 1/2$, but a BCDW  is stable at around $x
\sim 0.4$, as observed experimentally. The most interesting result
of our calculation is the following: in a large part of the phase
diagram an intermediate state actually turns out to be most
favorable. This state is a superposition of a BCDW (Zener polaron)
and a SCDW (of CE-type). As one can easily see, a superposition of
Figs.1a and 1b results in a state in which the two Mn ions in one
dimer are inequivalent.  This produces a dipole moment on each
pair, and the resulting state is therefore ferroelectric. The
appearance of  ferroelectricity in the case of coexistence of a
bond-centered CDW (Peierls state) and the site-centered CDW was
previously observed in some quasi one-dimensional compounds
\cite{Brazovskii}.  But, to the best of our knowledge, such a
state was never discussed in the context of the three-dimensional
TM oxides. In this respect it is noteworthy that first
measurements of the dielectric constant in these manganites show a
strong anomaly at the CO phase transition, which might be related
to ferroelectricity \cite{Simon}.

{\it Qualitative considerations.} It is well established that
charge and orbital structures are closely related to the magnetic
ordering  \cite{Goodenough}. In all the papers where either the
usual SCDW \cite{Wollan} or the BCDW (Zener polarons)
\cite{Carvajal} are considered, the magnetic structure is taken to
be of the CE type (ferromagnetic zig-zag chains stacked
antiferromagnetically, Fig.1a). However in the BCDW picture the
situation can change drastically due to orbital ordering existing
in the structure  of Fig.1b (predominantly $d_{x^2}$-orbitals on
horizontal dimers and $d_{y^2}$ on vertical ones, see below). Due
to the double exchange mechanism the coupling between spins inside
a Zener polaron is definitely strongly ferromagnetic. The
high-spin polaron dimers
---if we treat them as localized spins--- form a triangular
lattice, see Fig.1c, with a relatively strong antiferromagnetic
coupling between neighboring parallel dimers. A different magnetic
coupling can be expected when the dimers are are oriented
perpendicular to each other. If these couplings would all be
comparable in magnitude and antiferromagnetic, then the angle
between neighboring high spin polarons on the triangular lattice
would be 120$^{\circ}$ and a tripartite Jaffet-Kittel magnetic
structure would be formed, Fig.1c. However from the orbital
structure of Zener polaron described above one  can immediately
conclude that the perpendicular coupling is definitely weaker. We
therefore expect that the magnetic structure tends to an
"orthogonal" ($\perp$) ordered state with four different spin
sublattices as shown in Fig.1b. We should mention that the neutron
scattering data  for Pr$_{0.6}$Ca$_{0.4}$MnO$_{3}$ can be equally
well described both by the conventional magnetic structure of
Fig.1a and by the orthogonal structure of Fig.1b
\cite{Carvajalprivat} that we propose here. In the overdoped
manganites R$_{1-x}$Ca$_{x}$MnO$_{3}$, with $x = 2/3, \ 3/4$ a
$\perp$ magnetic structure was already observed \cite{Radaelli}.

{\it Calculated phase diagram.} In our calculations  we use a
tight binding approach for the band structure of the manganites,
where both double exchange and superexchange are incorporated.
This approach proved to be quite efficient for overdoped ($x>0.5$)
\cite{vandenBrink99a} and half-doped \cite{vandenBrink99b}
manganites. In the double-exchange framework with strong on-site
Hund's rule coupling the motion of $e_g$-electrons is largely
determined by the underlying magnetic structure. In this approach
the stability of various magnetic and orbital structures is
determined by the competition of the band energy of the
$e_g$-electrons, favoring  ferromagnetism, and superexchange
interaction $J$ between localized ($t_{2g}$) spins, favoring
antiferromagnetism. The double degeneracy of conduction electrons
significantly modifies the conventional picture and leads to the
stabilization of more complicated magnetic structures, besides the
simple ferromagnetic (F) or antiferromagnetic (G) ones
\cite{vandenBrink99a}.  We calculate the  groundstate energy for
different magnetic structures: G-, C-  and A-and CE-type
antiferromagnetic states, the ferromagnetic (F) phase, and, in
addition, for  two new phases (with orthogonal ($\perp$) and
120$^{\circ}$ Jaffet-Kittel type magnetic ordering of spin dimers)
that we expect to be relevant close to $x \sim 1/2$, based on the
arguments above.

The effective  hopping matrix elements $t_{ij}^{\alpha \beta}$
depend on the type of the orbitals $\alpha, \beta$ and relative
orientation of a pair $i$, $j$, so that $t_{i,j \parallel
x}^{x^2-y^2 , \,  x^2-y^2 } = 3/4 t $, $t_{i,j \parallel z}^{z^2 ,
\, z^2} = t$ and $t_{ij \parallel x}^{z^2, \,z^2} = 1/4 t$
\cite{orbitalnote}. Besides this, there is the usual dependence of
hopping on the angle $\theta$ between neighboring spins:
$\tilde{t}_{i,j}^{\alpha, \beta} = t_{ij}^{\alpha, \beta} \cos
\theta_{ij}/2$. The antiferromagnetic superexchange energy per
bond is $J \cos \theta_{ij}$.  With these expressions it is
straightforward to calculate the band dispersion and total energy
of different phases, cf.  \cite{vandenBrink99a}.
In Fig.2. we show the resulting phase diagram as  a function of
doping $x$ and antiferromagnetic coupling $J$. For $x \gtrsim 1/2$
we find mainly the magnetic CE, C, A and F phase to be stable, in
accordance with previous results \cite{vandenBrink99a,ref2}. For $
x \lesssim 0.5$ we also find a large region where the BCDW
magnetic 120$^{\circ}$ state is stable. Most interestingly, we
also find a finite region of the $\perp$ phase. This phase is
stable for doping concentration from around $x \sim 0.4$, which
agrees with the fact that at this concentration the Zener polaron
state was found \cite{Carvajal}.

\begin{figure}
\includegraphics[width=0.55\columnwidth]{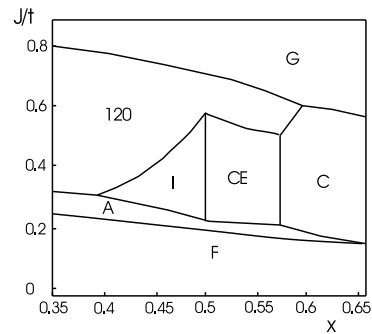}
\caption{Phase diagram of manganites for doping concentrations close to 1/2.
F: ferromagnetic, G: antiferromagnetic (AF) A:  F planes coupled AFM, C:
F chains coupled AF, CE: F zig-zag chains coupled AF, 120: Jaffet-Kittel
state and I: intermediate orthogonal phase. The electron density
is $n_{el}=1-x$.}
\end{figure}

From the calculation we know the corresponding wave functions  so
we can find the orbital occupation on a certain site by projecting
the wave function on that particular site. In agreement with the
standard picture we obtain that for the CE-structure the edge
sites  A, A' in Fig.1a. are occupied  by orbitals $d_{x^2}$ and
$d_{y^2}$, respectively. On the corner sites predominantly the
$d_{x^2-y^2}$ orbital is occupied with about 10 \% admixture of
$d_{z^2}$ density.

Let us denote the general orbital state  as $|\phi \rangle  = \cos
\phi/2 | z^2 \rangle + \sin \phi /2 | x^2 -y^2 \rangle$. For an
isolated dimer that points along the $x$ direction (the equivalent
D, D' sites in Fig.1.b) we then expect $\phi = 2 \pi /3$ for both
sites in the dimer, which corresponds to occupied $d_{x^2}$
orbitals on both sites. For an isolated dimer along the $y$
direction (C, C' sites in Fig.1.b.) we then expect $\phi = -2 \pi
/3$, corresponding to occupied $d_{y^2}$ orbitals. From the
calculation of the wave functions for the $\perp$ phase, where
there is an interaction between neighboring dimers, we find for
the $x$-dimer that $\phi \sim 130 ^{\circ}$, which is mainly a
$d_{x^2}$-state, as we had for a non-interaction dimer, but with
some admixture of $d_{x^2-y^2}$. Correspondingly, for the
$y$-dimers we find $\phi \sim -130 ^{\circ} $, which amounts to a
predominantly $d_{y^2}$-state with some $d_{x^2-y^2}$ character.
This admixture is important as it is driving the stability of the
$\perp$ phase. The superexchange tends to align the spins of
neighboring $x$-and $y$-dimers antiparallel. But in the $\perp$
phase these spins are only orthogonal, so that the hopping between
the dimers is possible, although reduced by a factor $1/\sqrt{2}$
from its maximum value. From this hopping between neighboring
dimers, which leads to the admixture of $d_{x^2-y^2}$ states into
the groundstate wavefunction, the system gains kinetic energy.
Note that the hopping amplitude between $d_{x^2}$ orbitals of
neighboring $x$-dimers (or between $d_{y^2}$ orbitals of
neighboring $y$-dimers, respectively) is much smaller, so that the
spins of neighboring dimers of the same kind tend to align
themselves fully antiparallel. Thus we conclude that the stability
of the $\perp$ state is caused by its orbital order that allows
for an optimal compromise between electron delocalization, causing
ferromagnetic double exchange, and antiferromagnetic superexchange
interactions.

{\it Charge ordering and ferroelectricity.} In the CE-phase the
edge and the corner sites are  crystalographically inequivalent.
This results in different total electron occupation numbers on
edge and corner sites. Already when the on-site Coulomb
interaction $U$ between electrons in different orbitals but on the
same site is included, the common picture of a checkerboard SCDW
arises in this system, with a valence Mn$^{3.5+\Delta q}$ of the
corner sites, Mn$^{3.5-\Delta q}$ of the edge sites and $\Delta q
\leq 0.1$ \cite{vandenBrink99b}. The CE-structure has a center of
inversion symmetry and, consequently, there are no electrical
dipoles moments present in the groundstate. This is also the case
in the pure $\perp$ structure as both Mn ions in each dimer are
equivalent \cite{ref1}. But there may exist an intermediate state
that does have  dipole moments: this intermediate state is a
superposition of the CE and $\perp$ phase. This state has a
magnetic structure in which the spins of $x$-dimers (spins within
each dimer are ferromagnetic) are antiparallel to one another, and
the $y$-dimers, also antiparallel between themselves,  are rotated
with a certain angle $\theta$ with respect to the $x$-dimers. The
angle $\theta$ characterizes intermediate phase as it determines
the amount of mixing between CE and $\perp$ that is present in the
groundstate: $\theta = 0$ corresponds to the CE-structure, while
$\theta = \pi/2$ corresponds to the $\perp$ one.

We find that the region of stability of this intermediate phase is
larger than that of a pure $\perp$-phase, and extends from the
concentrations $x \sim 0.4$ to  $x=0.5$; Fig.2 shows for which
values of $x$ and $J/t$ this mixed intermediate phase is stable.
The angle $\theta = 0$ for $x=1/2$, and it increases with
decreasing $x$, reaching the value $\theta = \pi/2$ ($\perp$
phase) at $x \sim 0.4$, see Fig.3. The intermediate phase is
therefore in general a superposition of the CE and $\perp$ phase,
except  near the phase boundaries in Fig.2. where the pure phases
are stable. In the intermediate phase there exist simultaneously a
dimer (Zener polaron) bond-centered CDW and  a checkerboard
site-centered CDW. This implies that in the intermediate state the
valence of the two Mn's in a dimer is different, inversion
symmetry is broken and each dimer attains a dipole moment. In this
phase the dipole moments of the $x$-dimers point e.g. to the right
in Fig.1 and all dipole moments of $y$-dimers point e.g. up
(depending on how precisely the inversion symmetry of the system
is broken globally) causing each $xy$-plane to acquire a net
ferroelectric moment.

We calculate the charge disproportionation  $\Delta q$ due to an
on-site Coulomb interaction $U$ between the electrons in the
Hartree-Fock approximation. In Fig.3. $\Delta q$ is shown as a
function of doping. Charge ordering starts at $x\sim 0.4$ where
the intermediate phase appears, and has its maximum at $x =1/2$,
which coincides with the transition to a pure CE-phase.

\begin{figure}
\includegraphics[width=0.5\columnwidth]{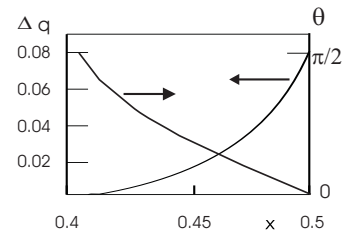}
\caption{Charge disproportionation  $\Delta q$ and mixing angle
$\phi$ (see text) for the $U=t$ in Hartree-Fock approximation in
the intermediate $\perp$ and CE phases.}
\end{figure}

It is experimentally observed that the CDW  deformations in
neighboring planes are in phase with each other. This implies that
there is a {\it net polarization} of the electric dipole moments
when the system is in the intermediate phase, which leads to a
ferroelectric ground state. Of course, in real materials
ferroelectric domains may appear. As also the conductivity of
these materials is not so low, it may not be easy to observe the
predicted ferroelectricity in CO manganites. But already first
measurements of the dielectric constant $\varepsilon$ do show a
strong peak in $\varepsilon(T)$ at the charge ordering temperature
\cite{Simon}. A more detailed experimental study of this
phenomenon would certainly be of interest in the present context.

A phenomenological symmetry analysis of possible  orderings in
R$_{1-x}$Ca$_{x}$MnO$_{3}$ was carried out in Ref. \cite{Plakida}.
The phases that we find fit exactly into the phenomenological
classification that these authors propose. They conclude that
three phases may exist: one with P2$_{1}$/m symmetry, which
coincides with our SCDW CE-structure; another of the Pmn2$_1$
type, corresponding to our BCDW orthogonal (Zener polaron) state,
and a structure with Pm-symmetry that corresponds to our
ferroelectric intermediate phase. Experimental structural
refinements do show that the actual symmetry of the
low-temperature phase of Pr$_{0.6}$Ca$_{0.4}$MnO$_{3}$ is Pm
\cite{Carvajal}, i.e. it is in the intermediate phase. Our
calculations provide the microscopic basis for this
phenomenological treatment and shows that, indeed, all three
phases are stable in a doping region close to $x \sim 0.4$.
Apparently, with increasing $x$ Pr$_{1-x}$Ca$_{x}$MnO$_3$ goes
from the $\perp$ or intermediate phase to the CE-phase at around
$x=1/2$, and then to the C-phase for overdoped samples. The
(LaCa)MnO$_{3}$-system is, due to a larger hopping $t$, apparently
below the $\perp$ phase in Fig.2: with increasing doping it goes
from F to CE  and then to the C-phase. In our phase diagram the
A-phase appears between the metallic F and "insulating" $\perp$
and CE-phases \cite{ref3}, whereas experimentally the A-phase
usually exists only for $x\gtrsim 0.5$ \cite{Kajimoto}; the reason
for this discrepancy is not clear at present.

We should remark that we only considered different homogeneous
states. It is known, however,  that in many correlated electron
models for the manganites the doped systems may be unstable with
respect to phase separation (PS) \cite{Dagotto}. Also in our
analysis we find for the intermediate phase a convex dependence of
the energy  on the electron density, i.e. there is the tendency
for PS into a pure CE and a pure $\perp$ phase with their
appropriate electron densities. However, the inclusion of the
long-range Coulomb interaction in model calculations of the
electronic properties prevents large-scale PS. Such a long-range
interaction has to be taken into account in the case of PS, in
order to guarantee the total charge neutrality of the system.
These Coulomb interactions allow only for the formation of weak
inhomogeneities with smooth boundaries, which implies that the
properties of the resulting PS state are similar to the properties
of the homogeneous state. Nevertheless, the question of what the
detailed properties of such an inhomogeneous state would be, is an
interesting problem and deserves further investigation.

{\it Conclusions.} We have shown that in manganites  with doping
concentrations close to $x=1/2$ a novel type of magnetic order can
appear besides the well known CE-type ordering. In this ($\perp$)
phase the neighboring spins are, depending on the direction of the
bond, either oriented parallel, antiparallel or perpendicular to
each other. Due to double exchange and orbital ordering, parallel
spins form strong dimer-like bonds, which we identify as the Zener
polaron state that was recently observed experimentally close to
$x=1/2$ \cite{Carvajal}. In considerable part of the phase diagram
an intermediate phase, which is a superposition of a CE-type and
orthogonal state, is stable. We predict that in this charge
ordered intermediate state inversion symmetry is spontaneously
broken and  ferroelectric moments appear. This may be the first
example of the appearance of ferroelectricity due to charge
ordering in transition metal oxides.

\begin{acknowledgments}
We are grateful to Z. Jirak, T. Fernandez-Diaz, P. Fulde, J. Hill,
M. Mostovoy, P. Radelli and J. Rodriquez-Carvahal for useful
discussion. This work was supported by the Netherlands Foundation
for the Fundamental Research of Matter (FOM) and by SFB608.
\end{acknowledgments}


\begin{references}
\bibitem{orbitalnote}  We denote the $3z^2-r^2$ orbital by $z^2$, $3y^2-r^2$ by $y^2$ etc.
\bibitem{ref2}  For $x>1/2$ one should in addition consider more complicated single and double-stripe structures.
\bibitem{ref1}  Strictly speaking this equivalence may be lost in the Zener polaron phase when GdGeO$_3$-type tilting distortions are taken into account.
\bibitem{ref3} Within our degenerate double exchange framework the phases for $x\neq 0.5$ are formally metallic;
the insulating character of all phases of (PrCa)MnO$_3$ is
expected to be due to disorder or phase separation -- the effects
that are not included in the present model.
\end{references}
\end{document}